\documentclass[twocolumn]{aastex631}
\usepackage{natbib, amsmath}
\usepackage{pifont}
\shorttitle{Feasibility of testing MBH seeding and growth using dwarf galaxies}
\shortauthors{Chadayammuri et al.}

\begin{document}

\title{Constraints from dwarf galaxies on black hole seeding and growth models \\ with current and future surveys}

\author{Urmila Chadayammuri}
\affiliation{Center for Astrophysics \ding{120} Harvard \& Smithsonian, 60 Garden St, Cambridge, MA - 02138, USA}

\author{\'Akos Bogd\'an}
\affiliation{Center for Astrophysics \ding{120} Harvard \& Smithsonian, 60 Garden St, Cambridge, MA - 02138, USA}

\author{Angelo Ricarte}
\affiliation{Center for Astrophysics \ding{120} Harvard \& Smithsonian, 60 Garden St, Cambridge, MA - 02138, USA}

\author{Priyamvada Natarajan}
\affiliation{Department of Astronomy, Yale University, 52 Hillhouse Ave, New Haven, CT - 06511, USA}

\begin{abstract}
Dwarf galaxies are promising test beds for constraining models of supermassive and intermediate-mass black holes (MBH) via their black hole occupation fraction (BHOF). Disentangling seeding from the confounding effects of mass assembly over a Hubble time is a challenging problem, that we tackle in this study with a suite of semi-analytical models (SAMs). We show how measured BHOF depends on the lowest black hole mass or AGN luminosity achieved by a survey. To tell seeding models apart, we need to detect or model all AGN brighter than $10^{37}\ \rm{erg \ s^{-1}}$ in galaxies of $M_* \sim 10^{8-10} \ \rm{M_{\odot}}$. Shallower surveys, like eRASS, cannot distinguish between seed models even with the compensation of a much larger survey volume. We show that the AMUSE survey, with its inference of the MBH population underlying the observed AGN, strongly favors heavy seed models, growing with either a power-law Eddington Ratio Distribution Function (ERDF) or one in which black hole accretion is tagged to the star-formation rate (AGN-MS). These two growth channels can then be distinguished by the AGN luminosity function at $> 10^{40}\ \rm{erg \ s^{-1}}$, with the AGN-MS model requiring more accretion than observed at z $\sim$ 0. Thus, current X-ray observations favour heavy seeds whose Eddington ratios follow a power-law distribution. The different models also predict different radio scaling relations, which we quantify using the fundamental plane of black hole activity. We close with recommendations for the design of upcoming multi-wavelength campaigns that can optimally detect MBHs in dwarf galaxies.
\end{abstract}
\section{Introduction} 
\label{sec:main}
Massive black holes (MBHs) are expected to reside in the core of every galaxy at $M_* \gtrsim 10^{10}M_\odot$, but their occupation fraction in galaxies at lower stellar masses are poorly constrained. Unlike stellar mass black holes, the origin of MBHs remains an open question, in particular, if there exist multiple pathways for their formation. Key candidate formation channels include remnants of Population III stars \citep{Bromm2003}, merged remnants in nuclear star clusters \citep{Portegies2004, Gurkan2004, Baumgardt2017, Natarajan2021}, the direct collapse of pre-galactic gas disks in which fragmentation and star formation is suppressed (DCBH) \citet{Begelman2006, LodatoPN2006, Mayer2010, Tanaka2014,Inayoshi2015}), or dark stars, powered by DM annihilation rather than nuclear fusion \citet{Spolyar2008, Freese2010}; see also the recent review by \citet{Volonteri2021}. Several seeding models are capable of reproducing observations in higher-mass galaxies under different assumptions for black hole growth \citep{Milosavljevic2009,Valiante2016,Ricarte&Natarajan2018b}, because observations largely probe later times in the Universe relative to the extremely early seed formation epochs. The models are all calibrated to match $z=0$ observations and reproduce observations of high black hole/galaxy masses as a function of time. However, to truly distinguish between the models, we either require observations at $z \gtrsim 7$ when seeds are expected to assemble  \citep{Treister2013,Ricarte&Natarajan2018a,Lusso2022} and/or lower mass galaxies at recent epochs \citep[see][for a review]{Reines2022}. 

Despite being modulated by complex growth mechanisms over cosmic time, some signatures of the initial MBH seeding mechanisms have been demonstrated to survive in late-time observables, primarily the low-redshift occupation fraction of MBH in dwarf galaxies \citep{Ricarte&Natarajan2018b, Volonteri2021}. While most, if not all, massive galaxies seem to host central MBHs with masses that are empirically seen to be correlated with host galaxy properties \citep{Kormendy2013, McConnell2013,Bogdan2015}, many low-mass galaxies may not host a central MBH at all; and if they do, may possibly follow different scaling relations \citep{Fontanot2015} or none at all \citep{Baldassare2020}. Multi-wavelength observations have been finding ever fainter MBHs in dwarf galaxies in the radio \citep{Reines2020}, via stellar dynamics \citep{Baldassare2015,Nguyen2018}, and from optical line ratios/variability \citep{Reines2013,Moran2014,Baldassare2018, Mezcua2020}. Any census of these low-mass MBHs is bound to be incomplete since some are too faint -- perhaps even completely inactive -- and therefore evade detection by these methods that rely on accretion activity. To infer the full occupation fraction of MBH, regardless of activity, one must incorporate detections as well as upper limits across multiple wavelengths and rely on some, preferably flexible models. Such work has been attempted successfully in the X-ray \citep{Miller2015, Gallo2019} and optical \citep{Kelly2013}.

But how to test MBH seed and growth models with the observed occupation fractions? The black holes that exist at $z = 0$, by and large, do not reflect the masses of initial seeds, as these are the result of growth averaged over cosmic time \citep{Mezcua2019}. We require models that trace the MBH population detected locally back to their high redshift seeds via growth histories. Semi-analytic models (SAMs) have proved to be a computationally efficient way to do so, as they involve populating and assembling halos over cosmic time, using well-defined physical models for black hole growth. In one such set of models, halos are populated with MBHs adopting various scaling relations, that permit tracking their growth over cosmic time in secular as well as merger-driven accretion modes \citep{Ricarte&Natarajan2018a, Ricarte&Natarajan2018b, Ricarte+2019}. SAMs enable fairly comprehensive probing of the parameter space of growth trajectories coupled with the different seeding mechanisms, quantifying how their combinations produce very different populations at $z = 0$. In this paper, we probe these differences in detail and examine which observables offer powerful discrimination, specifically on the issue of initial seeding. 

The paper is structured as follows. In \S 2, we present an overview of the semi-analytical model and the scaling relations used to create the mock catalogs. \S 3 compares the black hole occupation fraction using different cuts on black hole mass and luminosity and compares to X-ray observations. \S 4 makes recommendations for upcoming multi-wavelength surveys, discussing the capabilities and limitations of each and the synergies between them. We close with conclusions and future prospects in \S 5. 

\section{Tracking black hole growth over cosmic time}
\subsection{Semi-Analytic Models}
\label{sec:sams}
By construction, SAMs are set up to explore a range of seeding and growth modes for a population of black holes as a function of redshift. Time slices from these SAMs can then be compared to measured black holes. In this work, we focus on the SAM developed by \cite{Ricarte&Natarajan2018a} and explore new predictions for the low mass end of the local black hole mass function. In particular, we include the properties of the host galaxies explicitly in the scheme and do so flexibly by exploring multiple scaling relations between them and MBH properties. It is this new feature that allows us to delve more deeply into further predictions in combination with the Bayesian inference framework in this work. 

\begin{figure*}
    \centering
    \includegraphics[width=\textwidth]{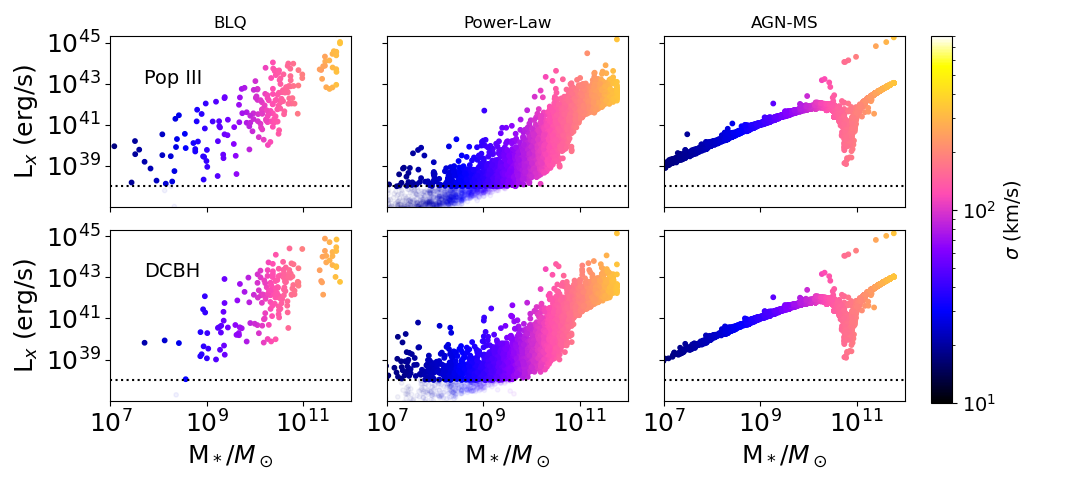}
    \caption{The $L_X-M_*$ scaling relation for each of the SAMs assuming $\epsilon=10\%$. The duty cycle $f_d$ marks the fraction of MBH with $L_X > 0$, and $\sigma$ is the root mean square error of the fit. The dotted line shows the flux limit of the AMUSE survey \citep{Gallo2008, Miller2012}. The power-law model predicts many AGN below this flux limit; the dashed lines, therefore, show an additional fit for only the AGN with $L_X > L_{lim}$. The color of each point represents the stellar velocity dispersion of the galaxy, as shown in the color bar.}
    \label{fig:sam-scaling}
\end{figure*}
Intriguingly, as first noted and discussed in detail in \citet{Ricarte&Natarajan2018b}, signatures of seeding survive to late times via the observed occupation fraction though there is dependence on the details of their mass assembly history. There are currently other SAMs from \citet{Somerville2008} \& \citet{Bower2010} and updated versions thereof, that do not focus on seeding and include recipes for star and galaxy formation as well as MBH growth, and these models inevitably have many more tuneable parameters. Here, we stick with a pared-down, phenomenologically driven model that is primarily focused on seeding signatures from nearly local observations ($z\sim 0$). The \citet{Ricarte&Natarajan2018b} SAM considers two seed models: either a massive and rare ``heavy'' seed typically produced from direct collapse following the seeding prescription of \citet{LodatoPN2006, LodatoPN2008}, or abundant ``light'' seeds drawn from a power law initial mass function between 30 and 100 solar masses motivated by cosmological simulations \cite{Hirano+2015}. 
For each seed, the SAM explored in \citet{Ricarte&Natarajan2018a} follows three different MBH growth prescriptions:
\begin{itemize}
    \item Power Law (PL) Model:  In this model, major halo mergers trigger a burst of growth at the Eddington rate until MBHs reach the mass determined by the observed local $M_\bullet-\sigma_*$ relation.  When MBHs are not in the burst mode, their accretion rates are drawn from a power law tuned to help reproduce low-redshift luminosity functions \citep{Ricarte&Natarajan2018a}.
    \item AGN Main Sequence (AGNMS) Model:  Major halo mergers trigger burst mode growth at the Eddington rate until MBHs reach the $M_\bullet-\sigma_*$ relation. When MBHs are growing secularly, their accretion rates are set to be $10^{-3} \ \times$ the star formation rate of their host. This mode is motivated by empirical relation determined by \citet{Mullaney+2012}. Since star formation is not actually tracked in this SAM, scaling relations are instead used to estimate stellar masses of galaxies as a function of halo mass and redshift \citep{Ricarte&Natarajan2018a}.
    \item Broad-line Quasars (BLQ) Model:  Major halo mergers (defined to be those with a halo mass ratio of at least 1:10) trigger burst mode growth. Additionally, at each time step the accretion rate is drawn from a distribution of Eddington rates observationally inferred for broad-line quasars in SDSS \citep{Kelly2013}.   Specifically, we adopt a redshift-evolving log-normal distribution fit by \citet{Tucci2017}\footnote{Here, the probability distribution of Eddington ratios $\lambda$ is given by $P(\lambda)=\exp(-(\ln\lambda-\ln\lambda_c(z))^2/(2\sigma(z)^2)/(2\pi \sigma(z)\lambda)$, where $\log \lambda_c = \mathrm{max}(\{-1.9 + 0.45z, \log 0.03\})$ and $\sigma(z) = \mathrm{max}(\{1.03 - 0.15z,0.6\})$}.  No additional steady accretion mode for mass growth is added since the AGN luminosity function is adequately predicted with this prescription \citep{Ricarte+2019}.  However, as discussed in this work, this assumption fails when considering deep surveys of dwarf galaxies that are not selected a priori to be AGN.
\end{itemize}
In this model, only 10$\%$ of the halo mergers lead to black hole mergers. This is done to prevent the overgrowth of BHs in the most massive halos, whose BHs would otherwise grow mostly from BH-BH mergers at low-redshift. We do not expect this assumption to have an impact on the dwarf galaxies considered in this work, which have much lower merger rates. In our study, we do not include satellite halos, since the SAM does not include a prescription for environmental processes such as ram-pressure/tidal stripping, which in turn is expected to affect the stellar mass as well as the growth of MBH. To increase our sample size, we compile the output from three snapshots at: $z = 0$, $z = 0.1$, and $z = 0.2$. This yields a total of 6,579 halos per seed $\times$ growth model, so the uncertainties from Poisson (small number) statistics on the predictions are small, even in the lower stellar mass bins. The scaling relations between galaxy and MBH properties evolve very marginally in this redshift range, and therefore no further modification is applied to the $z=0.1, 0.2$ halos.

\subsection{Creating mock catalogs from the SAM}
The SAMs provide the mass and mass accretion rates for MBHs at every snapshot. The simplest way to compute the luminosity is to convert the accretion rate into a bolometric luminosity with some efficiency $\epsilon_f$, and then use a bolometric correction factor to infer the luminosity in some specific energy/wavelength range. Like much of the literature, we assume the radiative efficiency of accretion $\epsilon_f$ = 0.1. 
The SAM is known to contain too little scatter \citep{Ricarte&Natarajan2018b} since the AGN are essentially always on. In practice, a duty cycle would increase the scatter in the observed instantaneous luminosity of the black holes. Meanwhile, we know that relatively complete surveys like AMUSE \citep{Gallo2008, Miller2012}, with a limiting luminosity of $2\times10^{38}  \ \rm{erg \ s^{-1}}$ at 0.5-7.0 keV, find AGN in $\sim 40\%$ of galaxies with $10^8 < M_*/M_\odot < 10^{12}$. AMUSE is a remarkably uniform and complete survey of AGN in nearby galaxies performed with the \textit{Chandra} X-ray telescope, observing 200 galaxies -- some in the field, some in the Virgo galaxy cluster -- to a limiting flux of $10^{38.3} \ \rm{erg \ s^{-1}}$, shown in Figure \ref{fig:sam-scaling} with a dotted line. While more recent catalogs exist, they are compiled from the \textit{Chandra} archive and are thus necessarily non-uniform in exposure \citep[e.g.][]{Gallo2019}. We measure the detected fraction of AGN in each SAM as:
\begin{align}
    w_i &= \frac{\Phi(M_{*,SAM})}{\Phi(M_{*,obs})}\\
    f_d &= \frac{\sum_{i=0}^{N_a} w_i }{\sum_{j=0}^{N_g} w_j}
\end{align}
where $w_i$ is the relative abundance of the host galaxy stellar mass in the SAM compared to the observations. The sum in the numerator is over the $N_a$ galaxies with $L_X > 0$ and in the denominator is over all $N_g$ galaxies. The weighting corrects for the different stellar mass functions of the SAMs and the observed sample. Both numerator and denominator include galaxies at $10^8 < M_*/M_\odot < 10^{12}$. The power-law and AGN-MS growth models predict the existence of many AGN below the AMUSE flux limit, shown as the dotted horizontal line in each panel.  

\begin{figure*}
    \centering
    \includegraphics[width=\textwidth]{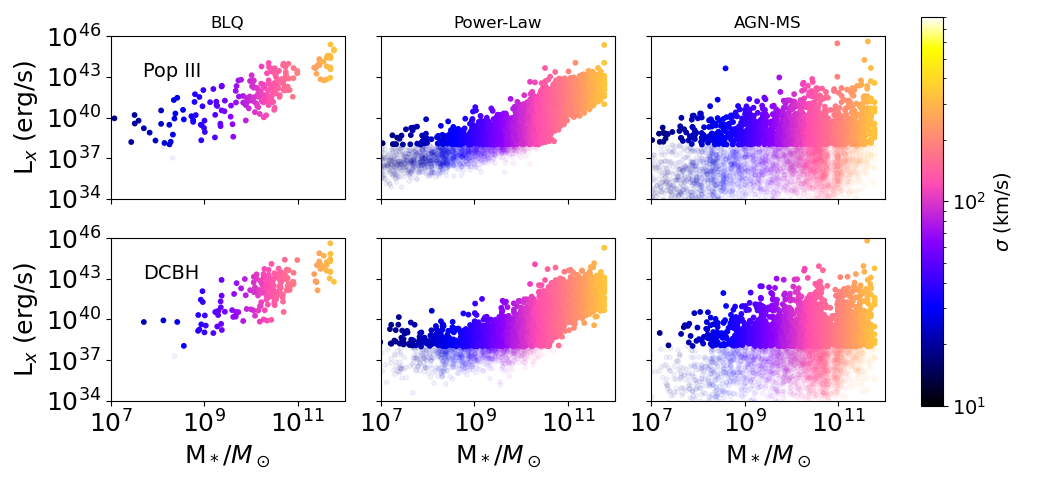}
    \caption{Same as Figure \ref{fig:sam-scaling}, but with enough scatter added to the $L_X$ to reproduce the observed AGN detection fraction of the AMUSE survey when each galaxy is weighted by the number density of observed galaxies of the same stellar mass. The additional scatter for each model is shown in Table \ref{tab:scalings}. This process was not carried out on the BLQ model, since its intrinsic duty fraction is far below what is observed.}
    \label{fig:plusscatter}
\end{figure*}

If the SAMs are taken at face value, 75-90$\%$ of the galaxies in the AGN-MS and PL models would have had AGN detections in an AMUSE-like survey. We, therefore, add scatter to the $L_X$ in these SAMs until the detected fraction, as defined above, is $40\%$ for an AMUSE-like survey. These values are shown in Table \ref{tab:scalings}, and the $L_X-M_*$ relations for the SAMs with the added scatter are shown in Figure \ref{fig:plusscatter}. 

\citet{Shen2020} measured the bolometric correction for the $0.5-2$~keV and $2-10$~keV X-ray bands and for the infrared B-band, as a function of bolometric luminosity. This corresponds to $1/k_{band} = L_{bol}/L_{band} =  10-30$ in each of the X-ray bands and $4-7$ in the B-band, each with a scatter of about 50$\%$. For each black hole, we draw from the bolometric correction from a normal distribution with the mean and scatter based on its bolometric luminosity.

\begin{figure*}
    \centering
    \includegraphics[width=\textwidth]{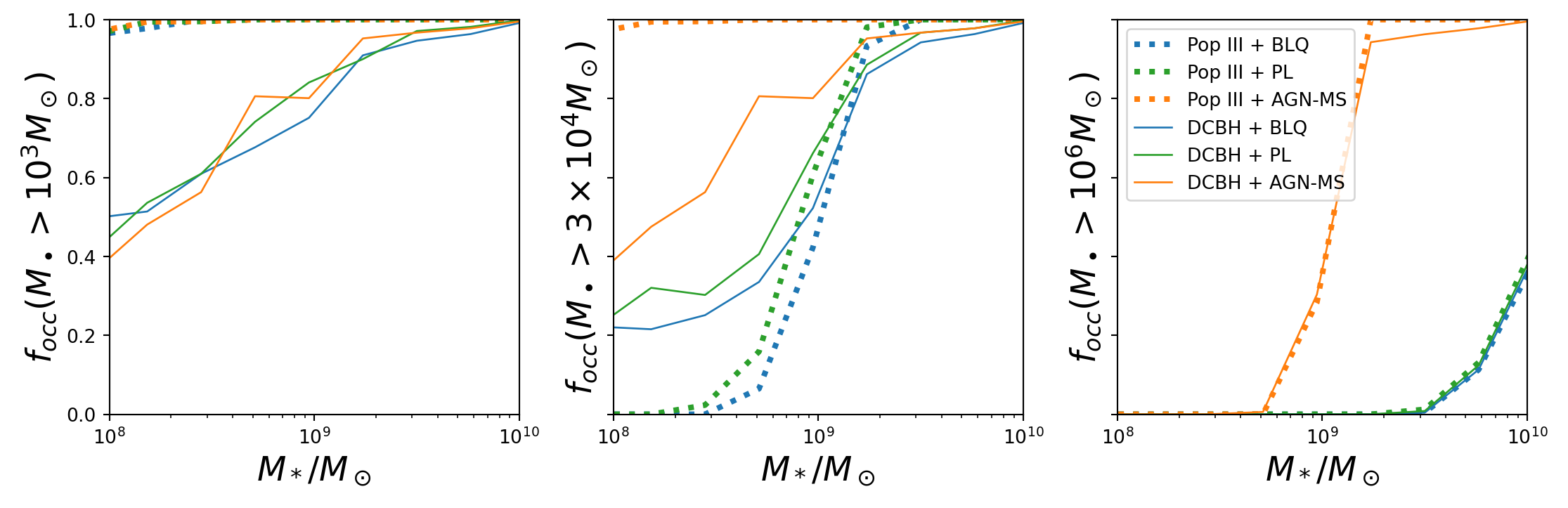}
    \caption{The fraction of galaxies containing black holes as a function of stellar mass that lie above different mass cuts. Dotted lines show the SAM with Pop III seeds, while solid lines are the heavier DCBH seeds. The colors indicate that the MBH grows following a BLQ (blue), PL (green), or AGN-MS (orange) model as described in \S \ref{sec:sams}.}
    \label{fig:focc_mbh}
\end{figure*}

\section{Results} 
\label{sec:results}
In this section, we compare the BHOF for the six SAM models (3 accretion modes $\times$ 2 seeding models) as detailed above with published values in the literature, applying cuts to the SAM outputs to match the observational selection functions. 

\subsection{Occupation fractions \& selection criteria}

We focus first on how observational selection can impact the derived BHOF and explore the optimal future survey strategies that will enable potentially discriminating between seeding models. We assume that black hole masses are available from the full range of possible observational probes and documented empirical correlations - either from dynamical modeling of galaxy cores or other techniques. AGN, for instance, have been identified via their optical variability \citep{Baldassare2018, Baldassare2020} or line ratio diagnostics \citep{Polimera2022, Cid2010, Kewley2013}; masses are then measured via reverberation mapping \citep[e.g.][]{Peterson2004} or direct dynamical modeling \citep[e.g.][]{Walsh2013,Seth2014, Nguyen2017}, and then used to construct scaling relations based on easily observable quantities like emission line strengths. Therefore, we proceed with the premise that we have in hand $M_{BH}$ measurements/upper limits for every single galaxy in our SAM survey sample. We first investigate the impact of the completeness of mass measurement on our inference of the BHOF.

Figure \ref{fig:focc_mbh} shows the occupation fraction for different values of black hole mass completeness. If we assume in the best-case scenario that all black holes above $10^3\ \rm{M_\odot}$ can be detected, then the Pop III seeds produce a higher occupation fraction than DCBH models. This is particularly noticeable for galaxies with $M_* < 10^9\ \rm{M_\odot}$. We note that this clearly reflects the efficiency and abundance of initial seeds, light seeds are more ubiquitous and are expected to form more efficiently compared to heavy seeds in the SAMs. If, however,  we can only detect BHs more massive than $3 \times10^4\ \rm{M_\odot}$, then light seeds growing through the AGN-MS  channel predicts $f_{\rm occ}=1$ for all stellar masses; for the other two growth channels, seed models become indistinguishable for $M_* < 4 \times10^8 \ \rm{M_\odot}$. If we can only find MBHs above $10^{6}\ \rm{M_\odot}$, information about seeding cannot be cleanly disentangled from accretion. These findings clearly show that the detectable present-day mass needs to be close to the initial assumed "heavy" seed mass from high redshifts, namely, $10^{4-5}\ \rm{M_\odot}$, in order to be disentangled from the confounding effects of the accretion physics. 

In summary, if the initial seeds have not grown much by any of the accretion modes or mergers, then their memory of seeding is retained in the BHOF as measured even at low redshift.
\begin{figure*}
    \centering
    \includegraphics[width=\textwidth]{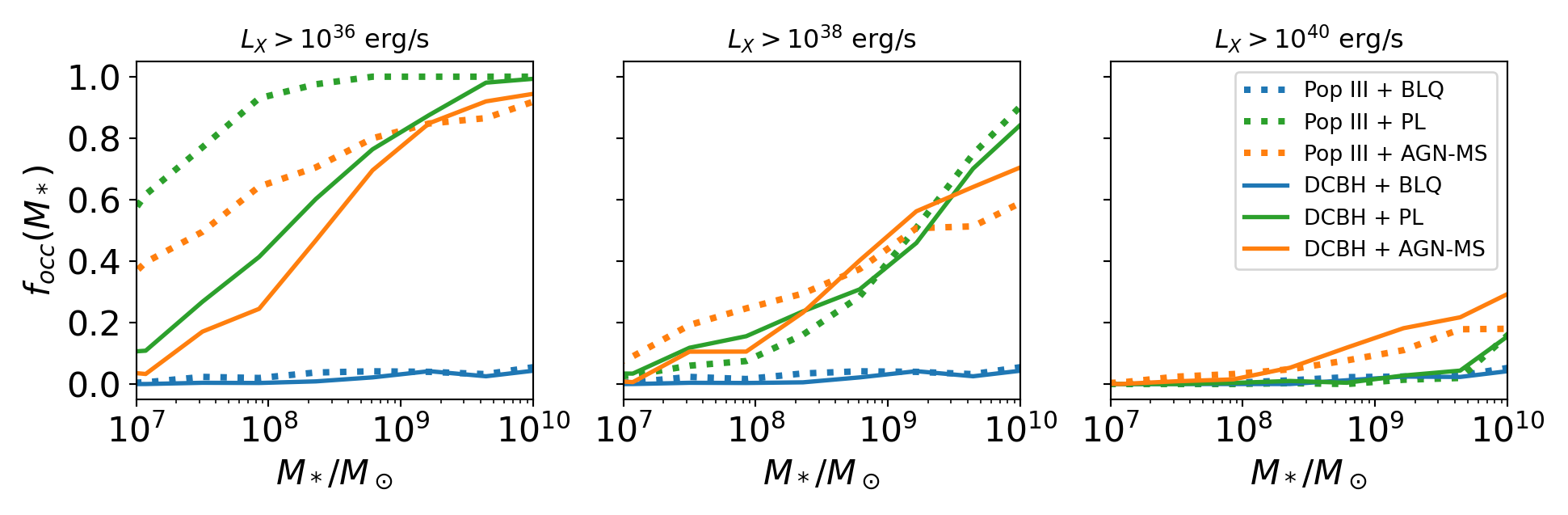}
    \caption{The detected AGN fraction for different survey luminosity limits. Here, the instantaneous accretion rate from the SAMs is converted to observed luminosity with the bolometric corrections from \citep{Shen2020} and a radiative efficiency $\epsilon_f$ = 0.1. The BLQ prescription assumes that MBH only accrete and radiate during a merger, and thus predict almost no activity at z=0. Since we do observe AGN at z=0, we can rule this model out. For $L_{\rm{lim}}  \leq 10^{37} \ \rm{erg \ s^{-1}}$, the Pop III and DCBH seed models predict very different occupation fractions at $M_* \lesssim 10^8\ \rm{M_\odot}$. Applying a higher luminosity cut then differentiates between the PL and AGN-MS models, with the latter predicting a higher luminous fraction.}
    \label{fig:sensitivity}
\end{figure*}

In practice, MBHs are usually detected when they are accreting and luminous. The scaling relations used to measure black hole masses regardless of wavelength deployed or method used, all suffer from large-intrinsic scatter \citep{Woo2002} and are most reliable at X-ray wavelengths where the obscuration is the lowest. Therefore, we stick to X-ray observations for comparison with SAMs for the rest of this study. Figure \ref{fig:sensitivity} shows the fraction of galaxies with AGN detected at $0.5-2.0$~keV as a function of the survey detection limit. The BLQ prescription is ruled out because it predicts almost no AGN in dwarf galaxies. A relatively shallow survey of $10^{40}\ \rm{erg \ s^{-1}}$ would only be able to marginally distinguish between the AGN-MS and PL growth channels modes, with the former predicting 10-20$\%$ more detections in dwarf galaxies; if all AGN above $10^{38} \ \rm{erg \ s^{-1}}$ are found, the detected fraction is identical for both these growth channels and both growth channels. We need a survey that either directly detects, or can model the presence of, all AGN with luminosities down to $10^{36} \ \rm{erg \ s^{-1}}$ if the heavy and light seed models are to be clearly distinguished.  

\begin{figure}
    \centering
    \includegraphics[width=0.47\textwidth]{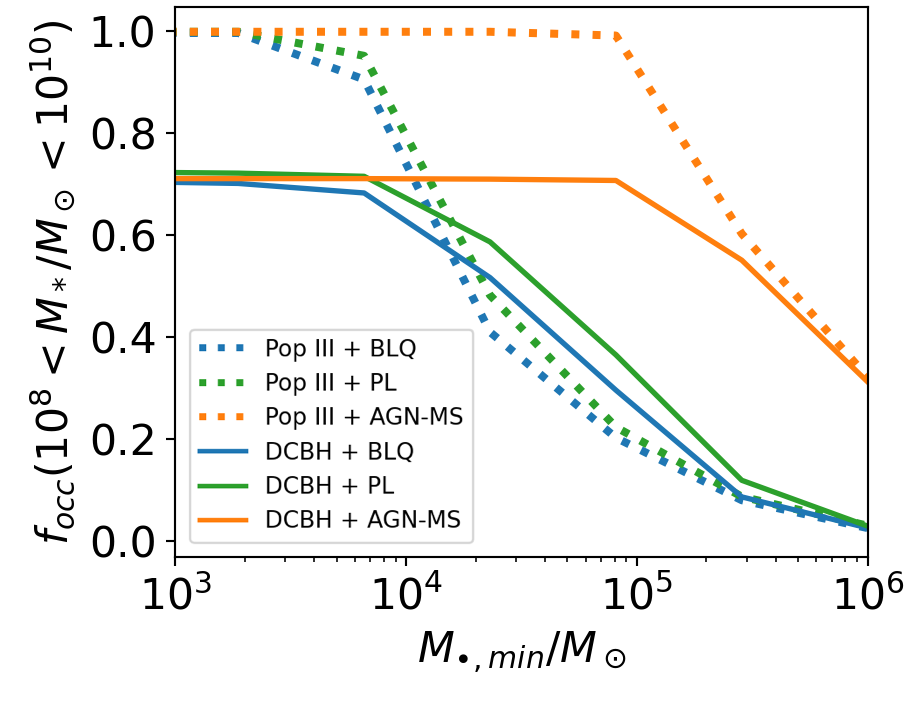}
    \caption{The total MBH occupation fraction for galaxies with $10^8 < M_*/M_\odot < 10^{10}$, as a function of the smallest detectable MBH mass. If the growth channel is AGN-MS, seed models can be distinguished for $M_{\bullet} > 10^5\ \rm{M_\odot}$; otherwise, all MBH above $10^4\ \rm{M_\odot}$ need to be detected for a given galaxy sample, assuming the sample is stellar mass complete.}
    \label{fig:minmbh}
\end{figure}

Figure \ref{fig:focc_mbh} and \ref{fig:sensitivity} reveal that most of the differences between the models are apparent at the lowest halo masses. Firstly, this highlights the importance of using a parent sample of galaxies that is stellar mass complete down to $10^8 \ \rm{M_\odot}$, pushing down an order of magnitude compared to current observational limits. Figure \ref{fig:minmbh} therefore shows the occupation fraction of galaxies with stellar masses in the range $10^8-10^{10} \ \rm{M_{\odot}}$ as a function of the minimum detectable black hole mass. If all the information we have is the black hole and stellar masses, we can only tell different seed models apart if we find every MBH at least as massive as $10^4M_\odot$. If we can separately confirm that the growth channel is like the AGN-MS model, then we can distinguish seeds using all MBH above $10^5M_\odot$. 

\begin{figure}
    \centering
    \includegraphics[width=0.47\textwidth]{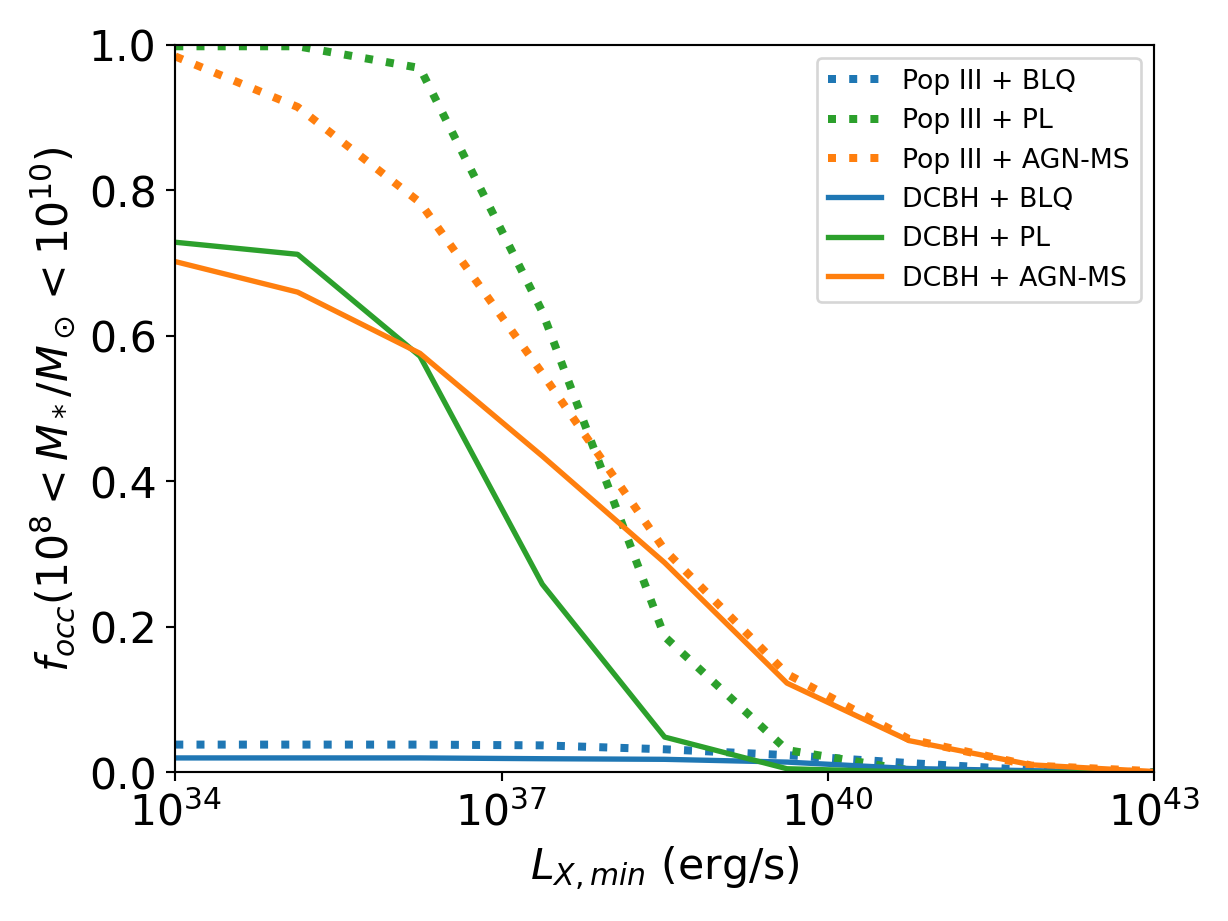}
    \caption{Similar to Figure \ref{fig:minmbh}, but for different limiting values for AGN luminosity at 0.5-2.0 keV detectable by an X-ray survey.}
    \label{fig:minLx}
\end{figure}

In practice, again, we almost always detect black holes when they are luminous. Figure \ref{fig:minLx} shows the fraction of galaxies with stellar mass $10^8 < M_*/M_\odot < 10^{10}$ with detectable AGN, as a function of survey limiting luminosity. Heavy and light seeds can be distinguished if the survey has a limiting luminosity better than $10^{37} \ \rm{erg \ s^{-1}}$. 

\begin{figure}
    \centering
    \includegraphics[width=0.47\textwidth]{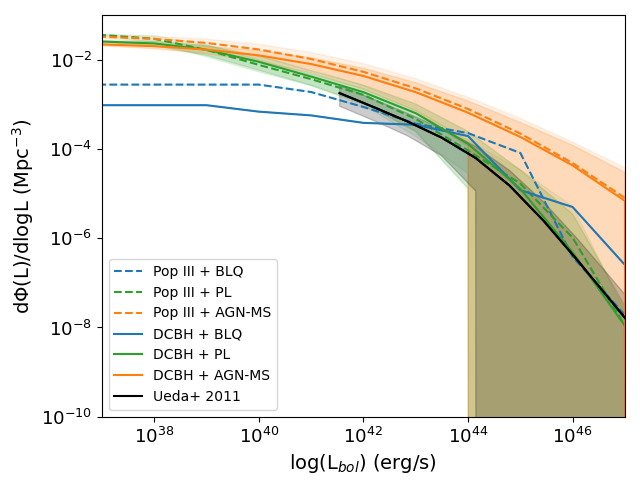}
    \caption{The bolometric luminosity function of the $z = 0$ AGN produced in the different SAMs. The three growth channels produce very distinct luminosity functions at z=0.}
    \label{fig:lumfunc}
\end{figure}

Figure \ref{fig:lumfunc} shows the AGN bolometric luminosity function for the six models, with observations from \citet{Ueda2011} in black for comparison. The uncertainties incorporate 1000 draws from the uncertainty in the bolometric corrections; the BLQ model uncertainties are not shown because, due to its low fraction of active MBHs, has error bars down to 0 at all luminosities. The AGN-MS and PL models have diverging luminosity functions above $10^{42} \ \rm{erg \ s^{-1}}$, with the former model requiring more growth at late times to match the $M_\bullet-\sigma$ relation. The PL model is thoroughly consistent with the observations, although the AGN-MS model cannot be entirely ruled out due to the large scatter required to bring it into an agreement with the observed active fraction in the AMUSE survey.  

\subsection{Coupling the SAMs to Bayesian inference}

As noted above, the BHOF depends on the minimum AGN luminosity that we can measure \textit{or model}. Studies like \citet{Miller2015} compute the BHOF from the observed AGN luminosities and upper limits, using a model where the $L_X-M_*$ relation contains log-normal scatter. Therefore, this BHOF accounts for AGN even below the nominal luminosity limit of the survey, which is $10^{38} \ \rm{erg \ s^{-1}}$. Since they did not publish the scaling relations inferred by their Bayesian model, we re-measure it here. To recap, the original study assumed that the BHOF had the functional form of a sigmoid function:
\begin{equation}
\label{eq:focc}    
f_{\rm occ} = \frac{1}{2} \left[1 + \tanh\left({2.5^{|8.9 - \log_{10}M_{*,0}|}\log_{10}\frac{M_*}{M_{*,0}}}\right)\right]
\end{equation}
The tuning parameter $M_{*,0}$ dictates how rapidly the occupation fraction jumps from 0 to 1; in other words, a higher value for this parameter means that galaxies with lower stellar masses are less likely to host SMBH than their more massive counterparts. Put another way, a higher $M_{*,0}$ means a lower BHOF.

A Bayesian inference model then constrains $M_{*,0}$ simultaneously with the normalisation, slope and scatter of the $L_X-M_*$ relation:
\begin{equation}
    \label{eq:lx-mstar}
    \log(L_X/\rm erg\cdot s^{-1}) = \alpha + \beta \times \log(M_*/M_\odot)
\end{equation}
This approach thus assumes 1) that the AGN duty cycle can be described in terms of the time spent below the detection limit if the $L_X-M_*$ scaling relation is a power law with log-normal scatter and 2) that we have no prior knowledge of this scaling relation. Again, Figure \ref{fig:plusscatter} shows the $L_X-M_*$ relation for each SAM, with points below the AMUSE detection threshold shown as translucent. 

Right away, we can rule out the BLQ prescription as the primary growth channel, because this predicts that only $3-4\%$ of the galaxies at $z < 0.2$ contain AGN, far below the $40\%$ detection fraction in AMUSE. This prescription was motivated from observations of the brightest quasars, the high luminosity end of the SDSS quasar LF. Since the abundance of bright quasars drops precipitously at low redshifts, the BLQ SAM is self-consistent in predicting low active fractions at low-z. The fact that this differs with observations of low-mass galaxies shows that this growth channel, which is bursty and merger-driven, cannot be important in dwarf galaxies. Next, we aim to distinguish between the AGN-MS and PL growth channels.

\begin{table}[]
    \centering
    \caption{Scaling relations for the PL and AGN-MS SAMs, measured for $10^8 < M_*/M_\odot < 10^{10}$ and with $\sigma_{add}$ adjusted so that $\sim40\%$ of the galaxies have AGN with $log(L_X/\rm{erg \ s^{-1}) > 38.3 }$, the limiting luminosity of the AMUSE survey.}
    \begin{tabular}{l|c|c|c|c}
        Model & $\sigma_{\rm add}$ & $\beta$ & $\alpha$ & $\sigma_{\rm intr}$ \\
        \hline
        PopIII - PL & 0.5 & 0.50 & 34.0 & 1.13\\
        PopIII - AGN-MS & 1.9 & 0.36 & 34.8 & 0.28 \\
        DCBH.- PL & 0.6 &  0.6 & 37.3 & 1.28 \\
        DCBH - AGN-MS & 1.8 &  0.35 & 34.7 & 0.29
    \end{tabular}
    \label{tab:scalings}
\end{table}

Both the power-law and AGN-MS, models predict very high (75-99$\%$) occupation fractions. From Figure \ref{fig:sensitivity}, we know that we need to apply a luminosity cut to the SAMs before comparing them to observations. For the \citet{Miller2015} study, this is not the same as the limiting luminosity of the observations. Rather, the BHOF in that study includes all the AGN whose luminosity falls within log-normal scatter from the mean value of the power-law $L_x-M_*$ relation. Therefore, we start by measuring this $\sigma$.

The probability of a parameter set {$\alpha, \beta, \sigma, M_{*,0}$} for detections is defined as:
\begin{equation}
\begin{split}
p(\log(L_X) | \log(M_*))  = f_{occ}(M_*, M_{*,0}) \times\\ 
N(\log(L_X) | \mu = (\alpha + \beta\times\log(M_*), \sigma = \sigma)
\end{split}
\end{equation}
where $N(x | \mu, \sigma)$ is the normal probability distribution function centered at $\mu$ with standard deviation $\sigma$. For a $n$-th non-detection, the probability that there is a black hole is given by the latent variable $I_n$, which in \citet{Miller2015} depended only on the scaling relation and occupation fraction. 
\begin{equation}
\begin{split}
    L_{X,\rm Pred} &= \alpha + \beta\times\log(M_*) \\
    \Phi &= C(L_{\rm lim} - L_{X, \rm pred}/\sigma)\\
    p_{BH}(L_{\rm lim}, M_*) &= \Phi/((1 - f_{occ}(M_*, M_{*,0}) + f_{occ}\times\Phi)
\end{split}
\end{equation}
where $C$ is the cumulative probability of the normal distribution. This says that a non-detection means either that there is no black hole, or there is a black hole and it is on but the luminosity is below the detection limit $L_{\rm lim}$. The latent variable $I_n$ is then set to either 1 or 0 by drawing from a binomial distribution with $p_{\rm true} = p_{BH}$. For further details of the inference pipeline, we refer the reader to \S 2 of \citet{Miller2015}.

\begin{figure}
    \centering
    \includegraphics[width=0.47\textwidth]{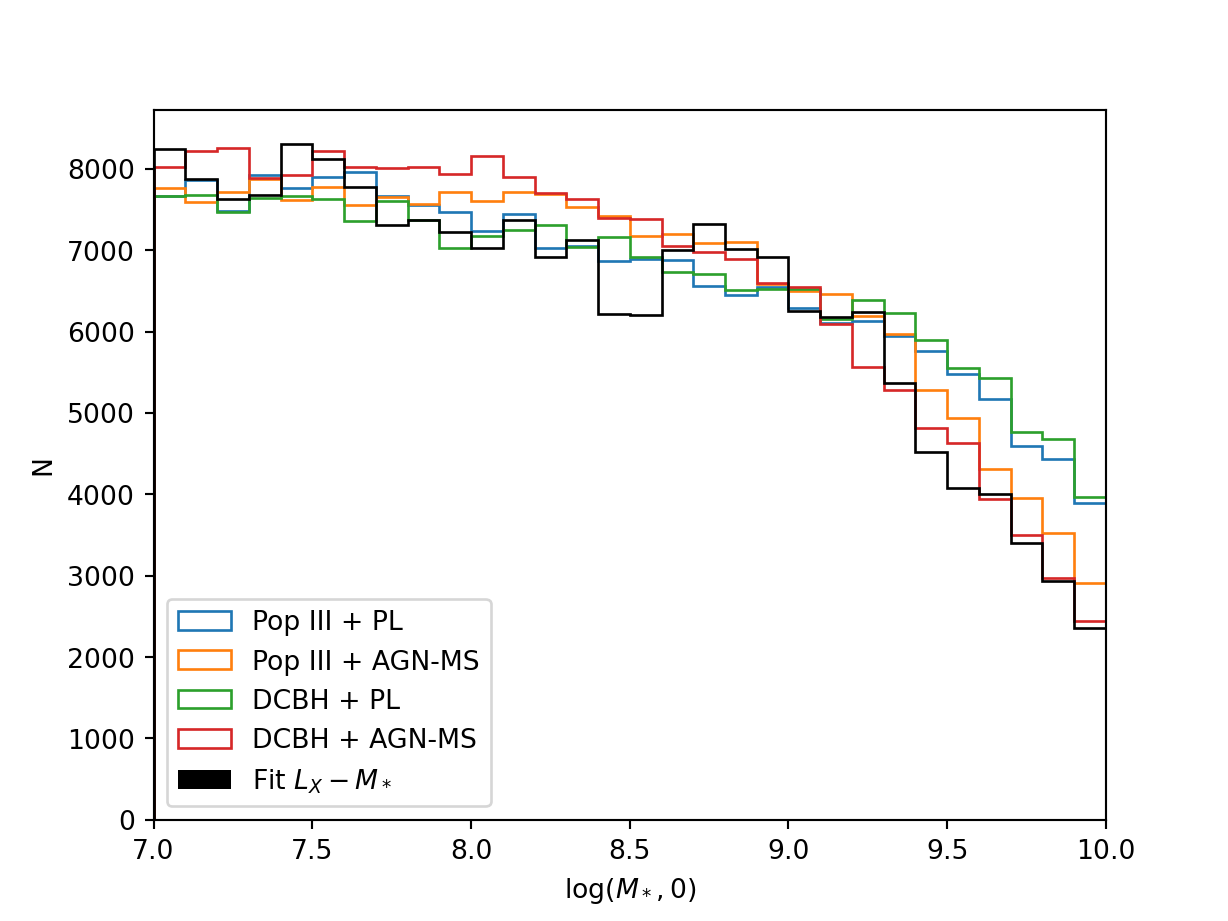}
    \caption{Posterior distribution on the tuning parameter $M_{*,0}$ of the BHOF. The black line allows the MCMC sampler to fit the parameters of the scaling relation in addition to $M_{*,0}$, whereas the remaining lines use the scaling relations from Table \ref{tab:scalings}. The power-law growth models have larger intrinsic scatter and are therefore slightly less constraining.}
    \label{fig:mcmc-test}
\end{figure}

As a test, we run the AMUSE sample through an MCMC pipeline four times, each time using a different scaling relation between $L_X-M_*$ in Table \ref{tab:scalings}, inferred from Figure \ref{fig:sam-scaling}. The resulting posterior on $M_{*,0}$, the stellar mass at which occupation fraction is $50\%$, is shown in Figure \ref{fig:mcmc-test}. Surprisingly, these posteriors are almost indistinguishable from each other. 

The best-fit value of $\sigma$ for AMUSE is 2; as seen in Table \ref{tab:scalings}, this is very comparable to the total scatter required in each SAM to match the AMUSE detection fraction. Thus, the BHOF inferred assuming an $L_X-M_*$ relation with such a large scatter would capture all the AGN in the SAMs with both the PL and AGN-MS growth channels, and no luminosity cut needs to be applied on the SAM MBHs.

\begin{figure}
    \centering
    \includegraphics[width=0.47\textwidth]{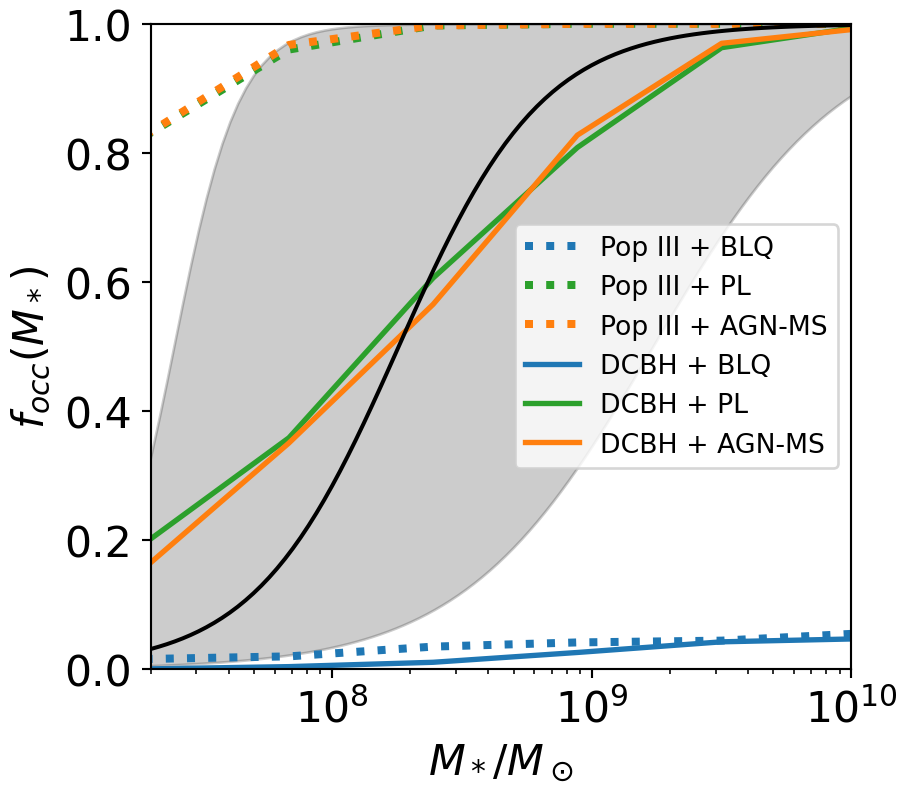}
    \caption{The AMUSE best fit and 1-$\sigma$ constraints on the BHOF, shown as the black solid line and gray shaded region. The SAM occupation fractions now use no luminosity cut, since the inference pipeline is designed to infer the presence of black holes below the detection threshold, as well. The observations clearly prefer heavy seed models.}
    \label{fig:comparison}
\end{figure}

Finally, Figure \ref{fig:comparison} compares the AMUSE constraints applied to each of the four SAMs. The observations are clearly consistent with the heavy seed models growing via either the PL or AGN-MS channels. The light seed models, even with large scatter in their $L_X-M_*$ relation, cannot be reconciled with an X-ray observed sample like AMUSE.

\begin{figure*}
    \centering
    \includegraphics[width=0.48\textwidth]{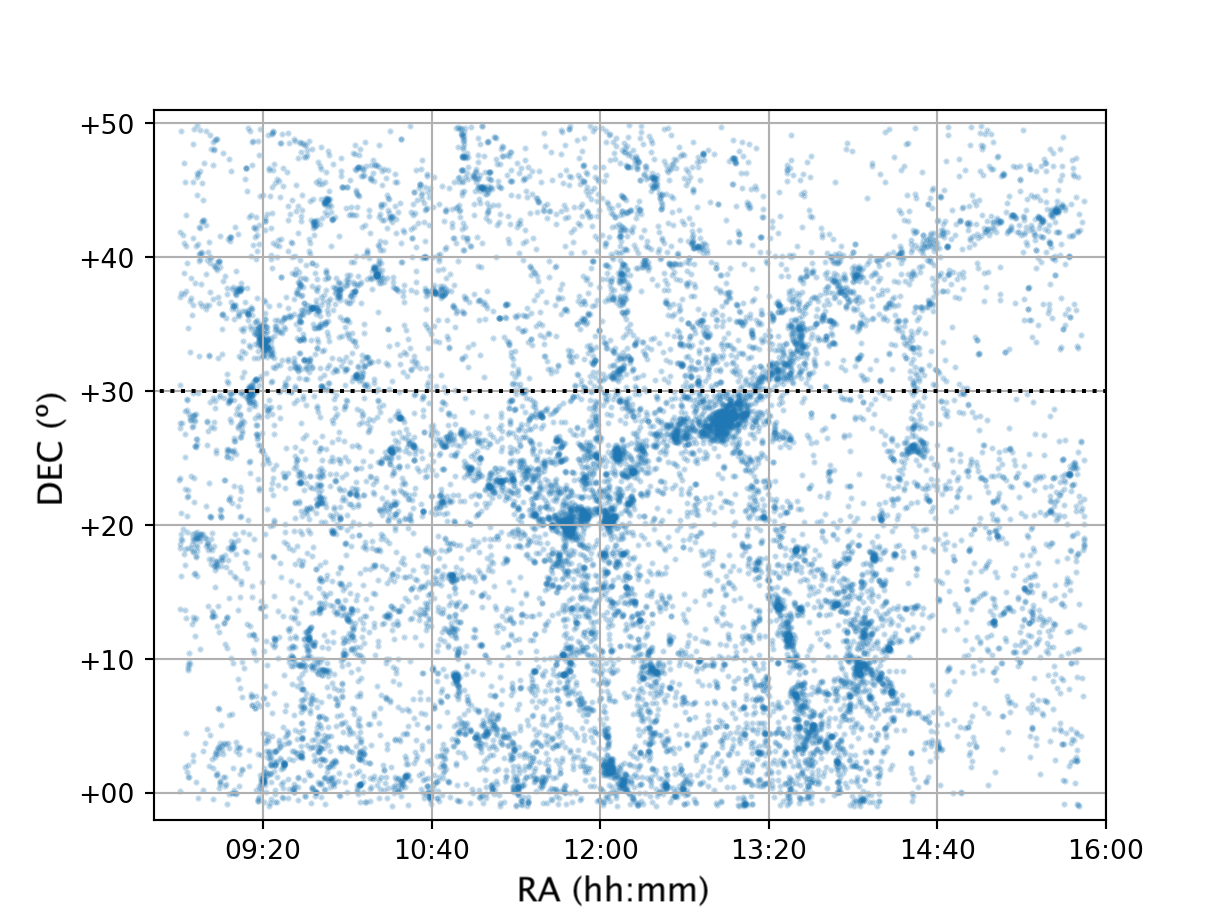}
    \includegraphics[width=0.48\textwidth]{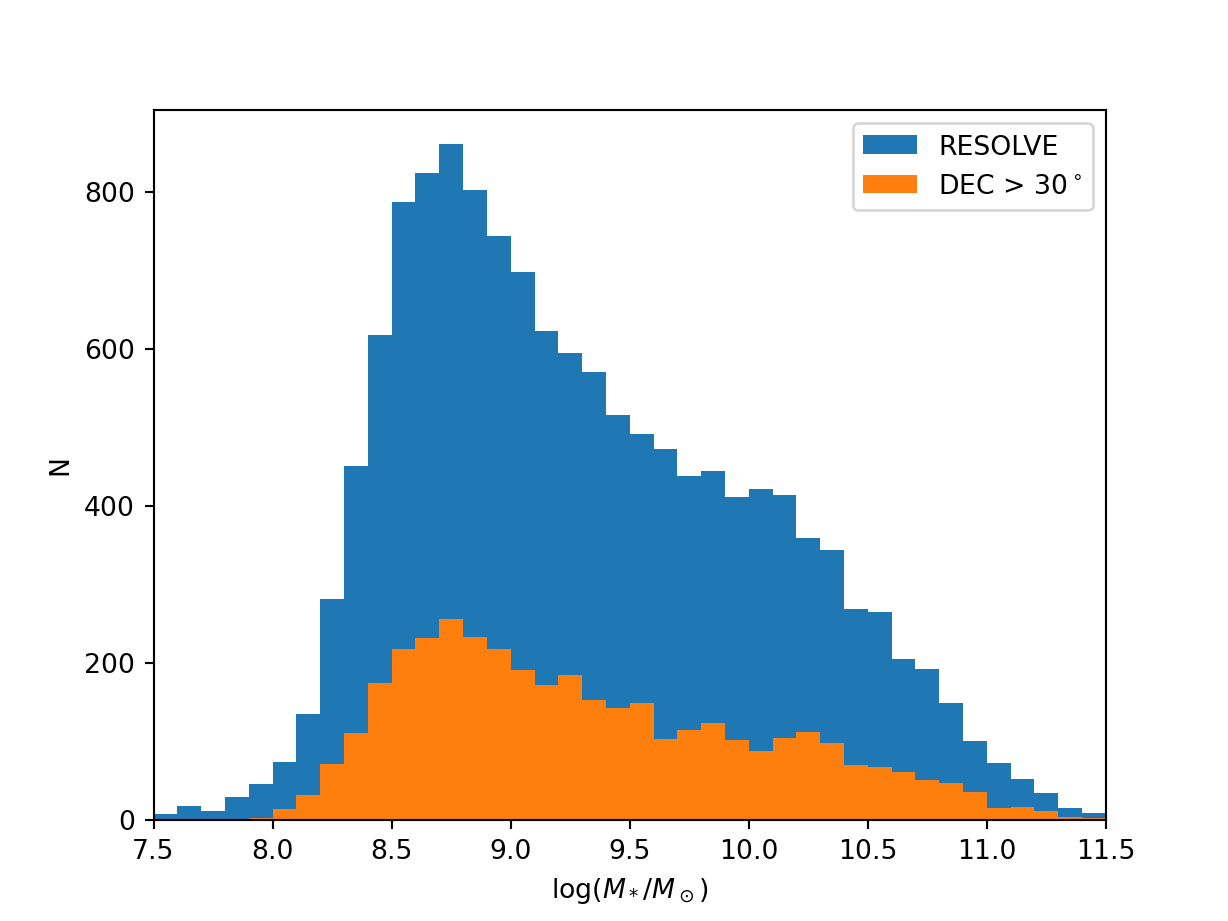}
    \caption{Map \textit{(left)} and stellar mass function \textit{(right)} of the RESOLVE survey \citep{Eckert2015}. This survey is stellar-mass complete above $5\times10^8\ \rm{M_\odot}$ out to z = 0.025, or a distance of $\sim110$ Mpc. The area above the dotted line in the map indicates the region where emission from the Milky Way is negligible, and extra-galactic surveys can probe fainter sources. The stellar mass function of the RESOLVE galaxies in this field is shown in orange on the right.}
    \label{fig:resolve}
\end{figure*}

\subsection{Prospects for future surveys}

A key result of our study is that distinguishing between MBH seeding and growth models relies crucially on detecting the least massive MBH (or least luminous AGN) in galaxy samples that are stellar mass complete down to at least $10^8\ \rm{M_\odot}$. We show here how such a survey could be constructed by following up on existing surveys.

The RESOLVE survey \citep{Eckert2015} was designed to be complete in baryonic mass down to $10^{9.3}\ \rm{M_\odot}$, and thus to even lower stellar masses, within $z < 0.025$ ($d < 110$ Mpc). Figure \ref{fig:resolve} shows the spatial (left) and stellar mass (right) distribution of the RESOLVE galaxies, with the latter demonstrating stellar mass completeness above $5\times10^8 \rm{M_\odot}$. The orange histogram shows only the galaxies above $DEC = 30^\circ$, where emission from the Milky Way is negligible, and even faint extra-galactic AGN could be detected. In principle, a survey could include galaxies at $|DEC| < 30^\circ$ as long as they are in the direction opposite to the Galactic Center; we are being conservative in our recommendation here to account for telescope scheduling difficulties, which may make it difficult to point at extragalactic fields at all times. The RESOLVE survey contains $>10,000$ such galaxies. 

\begin{figure*}
    \centering
    \includegraphics[width=\textwidth]{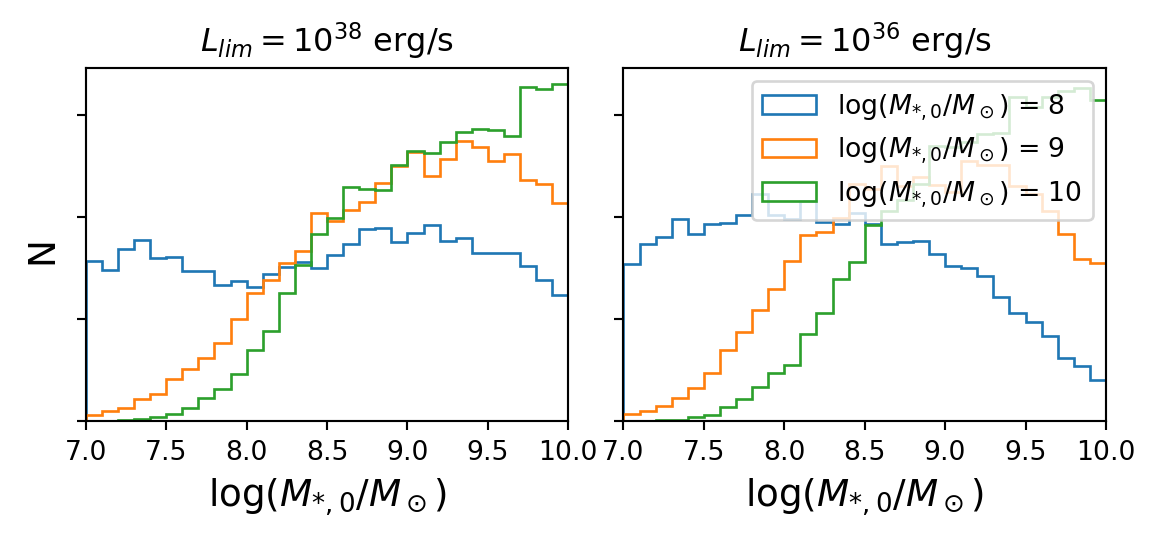}
    \caption{The recovered posterior on $\log(M_{*,0})$ for mock catalogs of 10K galaxies at z $<$ 0.025 observed with an X-ray survey with luminosity completeness above $10^{36}  \ \rm{erg \ s^{-1}}$. The faintest AGN created by the scaling relation is above this threshold, at $4\times10^{36} \ \rm{erg \ s^{-1}}$. The MCMC sampler assumes that the $L_X-M_*$ relation and its scatter are well-measured at higher stellar masses and do not change for dwarf galaxies. The posteriors are broadened by the scatter in the scaling relation and by the probabilistic nature of populating galaxies with black holes if $f_{occ}(M_{*,0}) < 1$. Nevertheless, the mode of the PDF tracks the true $M_{*,0}$ used to generate each mock.}
    \label{fig:mock}
\end{figure*}

Using this subset of galaxies as a parent sample, we create three different mock catalogs. Each assumes a different value of $M_{*,0}$ and thus a different BHOF. Where a MBH exists, we assign an X-ray luminosity drawn from the AMUSE observed $L_X-M_*$ relation, including the scatter. We pass it through the Bayesian inference pipeline described above, assuming survey luminosity limits of $10^{36}  \ \rm{erg \ s^{-1}}$ and $10^{38}  \ \rm{erg \ s^{-1}}$. Figure \ref{fig:mock} shows the posterior distributions recovered for each of the mocks. Higher occupation fractions are harder to rule out than lower ones, due to the large scatter inherent in the $L_X-M_*$ relation. While broad, the mode of each distribution matches the input value for $L_{lim} = 10^{36}  \ \rm{erg \ s^{-1}}$. For shallower surveys, if the intrinsic BHOF is high, the survey would return a flat posterior on $M_{*,0}$ even for this high level of completeness. On the one hand, this suggests that the uncertainties obtained from the AMUSE survey may be biased low. On the other hand, it confirms that higher values of $M_{*,0}$ would have been ruled out with high confidence.  

On the other hand, we have checked that the eFEDS survey, despite its uniform coverage overlapping with $\sim$500 times more galaxies than AMUSE and detection of several new AGN candidates in dwarf galaxies \citep{Latimer2021a}, yields flat posteriors on $M_{*,0}$, and that this situation is not expected to improve even with the deepest, polar regions of eRASS8. The key takeaway from this analysis is that survey depth, and the resulting completeness in stellar mass and MBH mass/luminosity, are much more powerful than survey volume when measuring the underlying BHOF. 

\section{Discussion}

\subsection{Confusion with stellar black hole systems}
A key challenge at low MBH masses, and therefore luminosities, is confusion with other sources such as Utra- or HyperLuminous X-ray sources (ULXs/HLXs) \citep{Swartz2004,Swartz2011}, which are likely the high Eddington-ratio end of the accreting stellar black hole population \citep{Stobbart2006,Sutton2013}, although some studies argue that the brightest of these may indeed be intermediate-mass black holes \citep{Miller2003,Bellovary2010,Sutton2012}. ULXs may be differentiated from IMBHs by a lack of radio emission or a position offset from the galaxy center \citep{Thygesen2023}, although \citet{Bellovary2021} showed that in the shallow potentials of dwarf galaxies, MBHs could live off-center. Multi-band X-ray observations can enable spectral studies to further differentiate between MBHs and high-accreting stellar black holes, and multi-epoch observations may find different variability scales. Further contamination at the faintest ends may arise from populations of X-ray binaries (XRBs), although this can potentially be modeled via low-scatter scaling relations with the stellar mass and star-formation rate \citep{Lehmer2010, Lehmer2019}.

\subsection{What would a low-redshift multi-wavelength campaign need to look like?}
Whether the black hole selection is by mass (Figure \ref{fig:focc_mbh}) or luminosity (Figure \ref{fig:sensitivity}), seeding models cannot be distinguished unless we find every MBH $\gtrsim 10^4\ \rm{M_\odot}$. This finding was also independently arrived at in recent work by \citet{Burke2022} using future variability surveys. These smallest MBHs have been recently discovered using optical variability \citep{Baldassare2018} and line diagnostics \citet{Baldassare2015, Mezcua2020, Polimera2022}. Emission line widths, such as $H-\alpha$, can yield black hole masses, although these can easily be confused with ongoing star formation. 

MBHs are, however, notorious for the wide variety and variability of their Spectral Energy Distributions (SEDs). This means that we do not yet have a complete understanding of how the emission from the reprocessed material feeding a black hole is distributed at different wavelengths. Below, we summarize some of the opportunities and challenges at various wavelengths for future low redshift surveys.

\subsubsection{Optical and infrared}

 Optical emission from AGN is highly susceptible to obscuration by dust, dust either in the torus of the AGN itself or elsewhere in the host galaxy \citep[e.g.][]{Barger2005}. Dust-absorbed optical photons are then re-emitted in the infrared (IR); however, infrared surveys thus far have lacked the sensitivity and/or spatial resolution to capture this population well enough \citep{Latimer2021b}. \cite{Marleau2017} found 62 AGN candidates in dwarf galaxies ($10^6 < M_*/M_\odot < 10^9$) in the WISE all-sky survey, but black hole masses require an assumption about Eddington ratios and therefore span three orders of magnitude. The current situation stands to be transformed by low-redshift data from the James Webb Space Telescope (JWST), particularly if there are appropriately designed surveys with well-understood selection functions.

\subsubsection{Spectroscopic line ratios}

A common way to find AGN is using spectroscopic line ratios, but these selection criteria were developed for more massive black holes and can fail for dwarf AGN. \citet{Polimera2022} found a population of AGN in dwarf galaxies using new line ratio diagnostics, exceeding previous detection rates in the galaxy sample by 3-16$\%$. \citet{Mezcua2020} found that for 23 of 34 AGN in dwarf galaxies with resolved IFU spectroscopy, the AGN signal gets washed out if integrated over the entire galaxy. Recently, \citet{Cann2018} has shown that more IMBHs should show up via coronal emission lines in JWST observations of nearby dwarfs. 

\subsubsection{Optical/IR variability selection}

Using a similarly phenomenologically motivated model, \cite{Burke2022} recently presented forecasts for the detection prospects of actively accreting black holes hosted in dwarf galaxies using their optical variability. Focusing on the intermediate black hole mass regime, they use observational constraints on optical variability as a function of black hole (BH) mass to generate mock light curves. In a similar vein to the work presented here, adopting several different models for the BH occupation function, including one for off-nuclear IMBHs, they quantify differences in the predicted properties of black hole mass and luminosity functions in local dwarf galaxies. Modeling the fraction of variable accreting black holes as a function of host galaxy stellar mass, including selection effects, they report that the BHOFs for ‘heavy’ and ‘light’ initial seeding scenarios can be discriminated with variability alone at the $2–3\sigma$ level for galaxy host stellar masses below $\sim 10^8\ \rm{M_\odot}$  with data from the upcoming planned synoptic surveys with the Vera C. Rubin Observatory. Current variability-centered searches for low mass BHs are starting to yield dividends, for instance, \citet{Ward2022} find $\sim$200 dwarf AGN using optical variability, and show that $81\%$ of them would be missed by spectroscopic searches.

\subsubsection{Radio wavelengths} 
AGN have also been detected in nearby dwarf galaxies through their radio emission, which stands out due to its very steep spectral index \citep{Yang2022, Sargent2022, Davis2022}. \citet{Liodakis2022} models the expected emission from jets in intermediate black mass holes in dwarf galaxies and predicts that $\sim40\%$ of them will be detectable with ngVLA, SKA and the Rubin observatory. 

\begin{figure*}
    \centering
    \includegraphics[width=\textwidth]{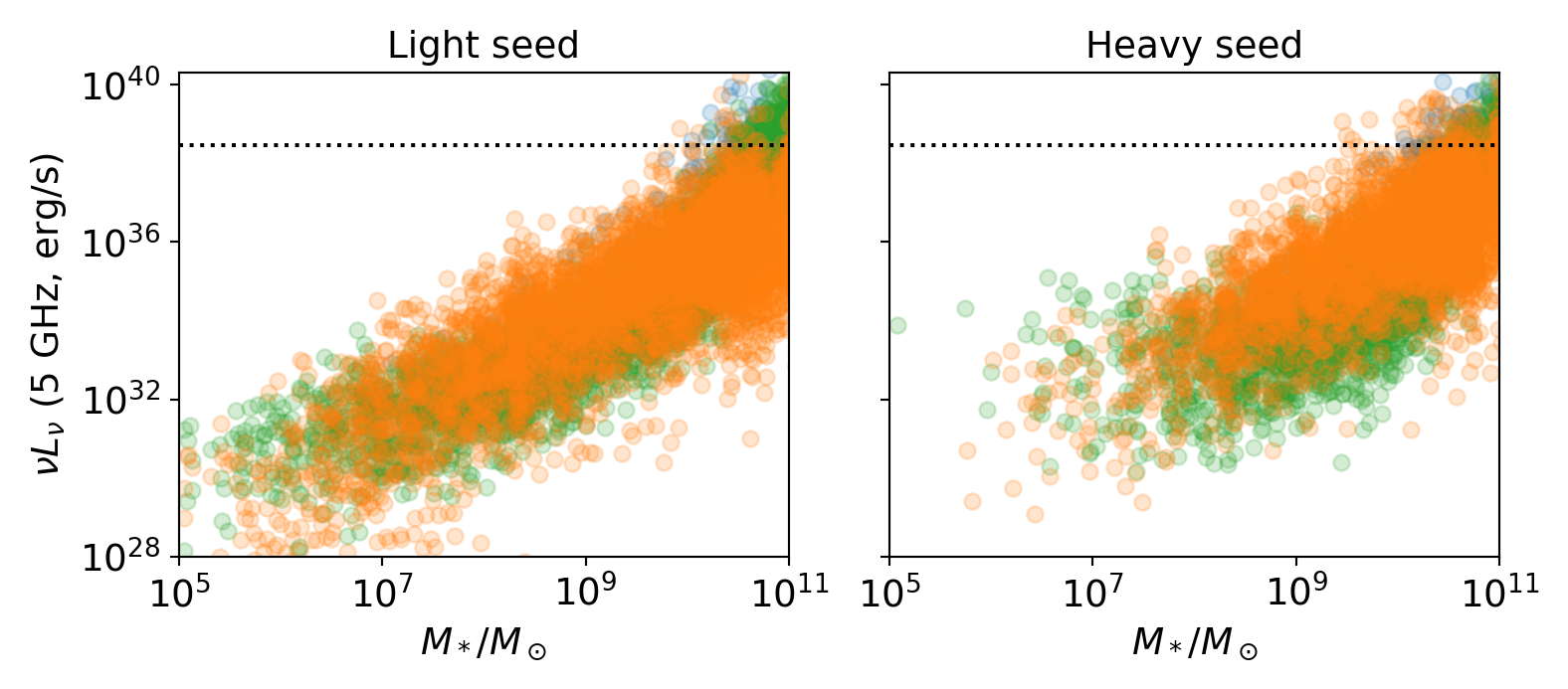}
    \caption{Predicted radio luminosity at 5 GHz from the three different growth channels, assuming the fundamental plane of black hole activity from \citet{Gultekin2019}. As with the X-ray, the AGN-MS model predicts higher radio luminosities on average and with lower scatter than the PL model, whereas the BLQ prescription predicts most SMBH to be inactive at z=0. The scalings are similar for the two seed models, but heavier seeds produce almost no sources for stellar masses below $10^7\ \rm{M_\odot}$. The dotted line shows a flux limit of 1 mJy, corresponding to the FIRST survey \citep{Becker1995}.}
    \label{fig:funplane}
\end{figure*}

Using the fundamental plane of black hole activity \citep{Merloni2003}, we can use black hole masses and X-ray luminosities to predict the expected radio emission for the different SAMs. The most recent measurement of this relation by \citet{Gultekin2019}:
\begin{equation}
\begin{split}
    log(M_\bullet/M_\odot) = \mu_0 + \xi_{\mu R}\log(L_R/10^{38}{\rm erg \cdot s}^{-1}) \\ + \xi_{\mu X}\log(L_X/10^{40}{\rm erg \cdot s} ^{-1})
\end{split}
\end{equation}
where the $L_R$ is the radio continuum luminosity at 5 GHz and $L_X$ is the X-ray luminosity integrated over 2-10 keV. The best fit to their observations was $\mu_0 = 0.55 \pm 0.22$, $\xi_{\mu R} = 1.09\pm0.10$ and $\xi_{\mu L} = -0.59\pm 0.15$, with an intrinsic scatter of -0.04 in log($M_\bullet$). Figure \ref{fig:funplane} shows the predictions derived using this observed fit for the expected $L_R$ for each of the SAMs. The heavy seed models predict slightly higher radio emission for dwarf AGN, but almost no sources in galaxies with $M_* < 10^7M_\odot$. Additionally, in the heavy seed scenario, the PL growth channel shows larger scatter in radio luminosities for the faint AGN in dwarf galaxies. This allows an additional way to distinguish between growth channels, other than the high-end luminosity function mentioned above, without having to assume that the same growth channels dominate for MBHs of all masses/luminosities. 

Combining optical, infrared, and X-ray observations would allow a self-consistent determination of the bolometric luminosity, removing the need for the uncertain bolometric correction factors that are currently adopted \citep{Thornton2008, Koliopanos2017}. Such a follow-up campaign to measure the masses of all the black holes detected via the range of methods above would greatly improve measurements of the BHOF.

\subsection{The possibility of multiple seeds}
In this idealised study, we have examined in detail the effects of seed and growth channels in isolation. It is likely that black holes may grow form via multiple channels. \citet{Spinoso2022} used a semi-analytic model to populate halos from the Millennium-II with both heavy and light seeds, and evolve their growth to z=0 using a two-phase prescription, corresponding to the observed quasar and radio modes of AGN. They find that with such a multiple-channel-seeding model, the slope of the black hole-stellar mass relation is flatter for dwarf galaxies than their more massive counterparts. We see this in Figure \ref{fig:sam-scaling} in the models that grow with a power law ERDF, respective of the seeding model, although the difference is starker for heavy seeds. Interestingly, this heavy-seed, power-law ERDF model is also the one that is most consistent with the AMUSE observations. This is a promising sign for future X-ray surveys, suggesting that we may find more dwarf AGN by simply pushing to slightly lower luminosities. 

\section{Conclusions}

We have assessed the feasibility of distinguishing between models of MBH seeding and growth in the presence of observational limitations. Using a suite of semi-analytic models, we applied cuts on the black hole mass, black hole X-ray luminosity and host galaxy stellar mass to mimic observational selection functions. We find that:
\begin{itemize}
    \item If we can find every MBH above $10^3\ \rm{M_\odot}$, light seeds predict a $\sim 100\%$ occupation fraction down to stellar masses of $10^8\ \rm{M_\odot}$, whereas for heavy seeds, this occupation fraction falls to $50\%$ at $10^8\ \rm{M_\odot}$. The total occupation fraction of MBHs at $10^8 < M_*/M_\odot<10^{10}$ is $\sim100\%$ and $\sim70\%$ for the light and heavy seeds, respectively.  
    \item If galaxies grow via the AGN-MS channel, the seed models can be distinguished if we can find every MBH above $3\times10^4\ \rm{M_\odot}$ and are stellar mass complete down to $10^8\ \rm{M_\odot}$. With a power-law ERDF, the occupation fraction of $>3\times10^4\ \rm{M_\odot}$ MBHs is 20$\%$ at $10^8-10^9\ \rm{M_\odot}$ for the heavy seed model, vs. $\sim0\%$ for a light seed. The two growth channels can then be told apart by looking only at black holes $>10^6\ \rm{M_\odot}$, which are much more abundant with the AGN-MS growth channel than the PL. 
    \item We use bolometric corrections from \citet{Shen2020} to determine X-ray luminosities for the simulated AGN. An X-ray survey of limiting luminosity $10^{36} \ \rm{erg \ s^{-1}}$  in the $0.5-2.0$~keV band can distinguish clearly between heavy and light seed models. If this limit is $10^{38}  \ \rm{erg \ s^{-1}}$, both seed models with either AGN-MS or PL growth channels are indistinguishable; at higher luminosity limits, most models produce $f_{occ} \sim 0$ at all stellar masses. 
    \item The BLQ prescription, which assumes purely merger-driven growth and reproduced the broad-line quasar luminosity function at high redshifts, predicts almost no luminous AGN at z=0, inconsistent with observations. We therefore conclude that it is not a viable mechanism for the growth of dwarf AGN.
    \item The constraining power of a flux-limited survey can be enhanced by physical modeling of the underlying MBH population. We use mock catalogs to confirm that Bayesian pipelines such as \citet{Miller2015} are able to recover the true intrinsic occupation fraction, using a combination of AGN detections and upper limits. The inferred BHOF is consistent with heavy seeds growing via either the PL or AGN-MS channels. 
    The two growth channels produce different luminosity functions at z = 0, however, with the PL model being more consistent with the observations of \citet{Ueda2011}.
    \item Using the fundamental plane of black hole activity, we predict that heavy seeds with produce more luminous dwarf AGN, but almost no sources in galaxies with $M_* < 10^7 \ \rm{M_\odot}$. While for light seeds the two growth channels produce identical radio signals, for heavy seeds, the PL model has a larger scatter in the $L_R-M_*$ relation. This will be testable with upcoming radio surveys like ASKAP and MeerKAT.
    \item We emphasise that future optical/infrared, variability-selected and radio surveys of AGN in dwarf galaxies must emphasise stellar mass completeness of the galaxy sample, since AGN non-detections are likely to be correlated with galaxy detectability ways that are non-trivial to model. 
\end{itemize}

In summary, we show concretely that it is possible to robustly test models of both the seeding and growth of supermassive black holes using low-redshift surveys. The surveys need to be carefully designed, achieving high completeness at low stellar and black hole masses and luminosities. Multi-wavelength campaigns will help break degeneracies. In most cases, pushing current surveys just one order of magnitude deeper will help distinguish between radically different early seeding models. 

\bigskip

\begin{small}
\noindent
\textit{Acknowledgements.} 
We thank Paul Tiede for helpful discussions of the statistical technique, and Ned Taylor for discussion on optical surveys. U.C. and \'A.B. acknowledge support from the Smithsonian Institution through NASA contract NAS8-03060. UC was also supported by NASA grant GO0-21070X. A.R. and P.N. gratefully acknowledge support from the Black Hole Initiative at Harvard University, which is funded by grants from the John Templeton Foundation and the Gordon and Betty Moore Foundation.
\end{small}

\bibliographystyle{aasjournal}
\bibliography{reference.bib}
\appendix
\section{Appendix}

\subsection{Bolometric corrections}

The most uncertain part of our model is the connection between the MBH mass and its X-ray luminosity. Observationally, we know that the MBH population is diverse - there are Type I \citep{Richards2006} and Type II AGN, Seyfert galaxies, FR I and FR II jets, blazars \citep{Giommi2009,Bottcher2017}, broad-line quasars (BLQs), narrow-line quasars (NLQs), low-luminosity AGN \citep{Almeida2021} and a variety of radio sources. A large part of this diversity has to do with dust obscuration, which affects X-ray wavelengths far less than optical/infrared (IR). However, there is expected to be intrinsic variability in the spectral energy distributions (SED) at different black hole accretion rates, depending on what process is driving the luminosity on (black hole event) horizon scales. 

We adopted the bolometric correction to the $0.5-2.0$~keV band from \citet{Shen2020}, which is a function of the bolometric luminosity; they found a similar scaling for the $2-10$~keV band. Their measurement agrees very well with \citet{Duras2020}; for the earlier \citet{Lusso2012} the corrections agree at $L_{bol} \gtrsim 10^{43} \ \rm{erg \ s^{-1}}$, but diverge significantly for fainter sources. Using \citet{Lusso2012} would thus yield much more pessimistic predictions for the X-ray surveys; we note, however, that these scaling relations were computed for AGN with $L_{bol} \gtrsim 10^{43}$ erg/s, far above the dwarf AGN regime. \citet{Vasudevan2007} found that the bolometric correction to the $2-10$~keV band is better correlated to the Eddington ratio. \citet{Woo2002} went so far as to argue that there was no correlation between black hole luminosity and mass other than due to circular reasoning in inferring the former from the latter. A relationship between the two is crucial for inferring black hole demographics unless we rely on dynamical measurements for the black hole mass, which is only possible for the nearest galaxies, or reverberation mapping, which is possible only for Type I AGN with an unobscured broad-line region. 

\subsection{Cosmic downsizing}

We do not know whether the Eddington distribution - the ratio of a black holes luminosity to the Eddington luminosity - is the same as for more massive galaxies, or if there is a ``downsizing" effect where less massive galaxies host, on average, more active black holes. There is then a degeneracy between the $M_*-L_{BH, X}$ scaling relation and the occupation fraction. The most robust constraints must jointly model these two effects, using an unbiased sample of host galaxies \citep{gallo2010}.

 \end{document}